\documentclass[iop]{emulateapj}

\catcode`\@=11
\def\gsim{\ifmmode{\mathrel{\mathpalette\@versim>}}
    \else{$\mathrel{\mathpalette\@versim>}$}\fi}
\def\lsim{\ifmmode{\mathrel{\mathpalette\@versim<}}
    \else{$\mathrel{\mathpalette\@versim<}$}\fi}
\def\@versim#1#2{\lower 2.9truept \vbox{\baselineskip 0pt \lineskip
    0.5truept \ialign{$\m@th#1\hfil##\hfil$\crcr#2\crcr\sim\crcr}}}
\catcode`\@=12

\shortauthors{S. Pellegrini}
\shorttitle{The temperature of hot gas halos of early-type galaxies}
\begin{document}

\title{The temperature of hot gas halos of early-type galaxies}

   \author{S. Pellegrini\footnote{E-mail:  silvia.pellegrini@unibo.it}}
\affil{Astronomy Department, University of Bologna, 
                       via Ranzani 1, 40127 Bologna, Italy   \\
email: silvia.pellegrini@unibo.it}

\begin{abstract} 
  Recently, the temperature $T$ and luminosity $L_X$ of the hot gas
  halos of early type galaxies have been derived with unprecedented
  accuracy from $Chandra$ data, for a sample of 30 galaxies covering a
  wider range of galactic luminosity (and central velocity dispersion
  $\sigma_c$) than before. This work investigates the origin of the
  observed temperatures, by examining the relationship between them
  and the galaxy structure, the gas heating due to Type Ia supernovae
  (SNIa's) and the gravitational potential, and the dynamical status
  of the gas flow. In galaxies with $\sigma_c\lsim 200$ km s$^{-1}$,
  the $T$'s are close to a fiducial average temperature for the gas
  when in outflow; at 200$<\sigma_c $(km s$^{-1})<250$, the $T$'s are
  generally lower than this, and unrelated with $\sigma _c$, which
  requires a more complex gas flow status; at larger $\sigma_c$, the
  $T$'s may increase as $\sigma_c^2$, as expected for infall heating,
  though heating from SNIa's, independent of $\sigma_c$, should be
  dominant.
All observed $T$'s are larger than the virial temperature, by up to
$\sim 0.5$ keV. This additional heating can be provided in the X-ray
brightest galaxies by SNIa's and infall heating, with a SNIa's energy
input even lower than in standard assumptions; in the X-ray fainter
ones it can be provided by SNIa's, whose energy input would be
required close to the full standard value at the largest $\sigma_c$. This
same energy input, though, would produce temperatures larger than
  observed at low $\sigma_c$, if entirely thermalized.
The values of the observed $T$'s increase from outflows to inflows;
the gas is relatively hotter in outflows, though, if the $T$'s are
rescaled by the virial temperature.  For $200<\sigma_c($km
s$^{-1})<250$, lower $L_X$ values tend to correspond to lower $T$'s,
which deserves further investigation.
  
\end{abstract}


\section{Introduction}\label{intro} 

The advent of the $Chandra$ X-ray observatory, with its unprecedented
sub-arcsecond resolution, allowed to study better than ever before the
main contributors to the total X-ray emission of early-type galaxies
(hereafter ETGs): the low-mass X-ray binaries (LMXBs; Fabbiano 2006),
a population of weak sources as late type
stellar coronae, cataclismic variables, and coronally active binaries
(Pellegrini \& Fabbiano 1994, Revnivtsev et al. 2008), the nuclear
emission due to a supermassive black hole (MBH; e.g., Gallo et
al. 2010, Pellegrini 2010), and a hot interstellar medium (ISM) with a
temperature of a few million degrees.  After careful subtraction
of the stellar (resolved and unresolved) and nuclear emissions, the
properties of the hot ISM could be characterized with unprecedented
accuracy.  Recently, this has been done for a sample of 30 normal
(non-cD) ETGs observed with $Chandra$ to a depth ensuring the
detection of bright LMXBs (Boroson et al. 2011, hereafter BKF).  This
is the first X-ray sample of ETGs covering a wide range of galactic
luminosity, central velocity dispersion $\sigma_c$, and hot gas
emission $L_X$, and with the X-ray properties of the hot gas (e.g.,
luminosity $L_X$ and average temperature $T$) derived in a homogeneous
way, using a complete and accurate procedure to subtract all kinds of
non-gaseous emission (nucleus, detected and undetected LMXBs, and
unresolved weak stellar sources).  This approach resulted in a larger
fraction of hot gas-poor galaxies than in previous samples, with $L_X$
extending down to much lower values than before ($\sim 10^{38}$ erg
s$^{-1}$), and showing a variation of up to $\sim 3$ orders of magnitude at
the same galactic luminosity (see also David et al. 2006, Diehl \&
Statler 2007, Memola et
al. 2009). Such a wide variation, even larger than previously found,
had been linked to the origin and evolution of the hot ISM, and had
provided evidence for the effectiveness of an internal heating
mechanism (as from type Ia
supernovae, hereafter SNIa's) to regulate the gas evolution and produce its
very different content in ETGs at the present epoch (Loewenstein \& Mathews
1987,  David et al. 1990, Ciotti et
al. 1991); the action of external agents (gas stripping, confinement,
accretion) to reduce or enhance the gas content was also invoked
(e.g., White \& Sarazin 1991, Brown \& Bregman 2000, Sun et al. 2007).

With this new characterization of the hot gas, BKF revisited the
relationships between fundamental properties of the hot gas and of the
host galaxy, as the $L_X-T$, $L_X-\sigma_c$ and $T-\sigma_c$
relations, where $\sigma_c$ is a representative measure of the depth
of the galactic potential well (Eskridge et al. 1995, O'Sullivan et
al. 2001, 2003).  $L_X$ correlates positively with $T$ and $\sigma_c$,
though with a wide variation at fixed $\sigma _c$ and
$T$. Interestingly, the best fit relation $L_X\propto T^{4.5}$, close
to what already known for X-ray luminous ETGs (O'Sullivan et
al. 2003), is still moderately strong among ETGs with low $T$ and
$L_X$; also, in the $L_X-\sigma_c$ relation, ETGs with $kT>0.4$ keV
are the X-ray brightest (with one exception), while those with
$kT<0.3$ keV are the X-ray faintest.  The least gas rich ETGs are then
the coolest ones, which seemed contrary to expectation, if low $L_X$
ETGs loose their ISM in an outflow (e.g., David et al. 1990, Ciotti et
al. 1991), and the hotter the gas, the stronger is the outflow (BKF).
On average $T$ increases with $\sigma_c$, and most ETGs lie above a
rough estimate of the gas virial temperature ($T_{\sigma}=\mu m_p
\sigma_c^2/k$), suggesting the presence of additional heating.  ETGs
with a moderate to high gas content ($L_X>5\times 10^{39}$ erg
s$^{-1}$) follow a trend roughly parallel to that of $T_{\sigma}$;
instead, ETGs with little hot gas ($L_X< 5 \times 10^{39}$ erg
s$^{-1}$) have a similar temperature for $\sigma_c$ ranging from 160
to 250 km s$^{-1}$.  This lack of correlation was attributed to a
different dynamical state of the hot ISM in gas-poor with respect to
gas-rich ETGs, though a full explanation of this aspect remained to be
found (BKF).

This work takes advantage of the new accurate measurements of the hot
gas properties, and of the fundamental relations $L_X-\sigma_c$ and
$T-\sigma_c$, derived down to galaxy masses and X-ray luminosities
smaller than ever before (BKF), to investigate the relationship
between $T$, the galaxy structure, the internal gas heating mechanisms
(SNIa's, and those linked to the gravitational potential), and the
dynamical status of the gas flow. To this purpose, a few
characteristic temperatures are introduced, depending on the nature of
the gas heating sources and the galaxy structure, and relevant for the
various gas flow phases; these characteristic temperatures are then
compared with the observed $T$ values.  In doing so, galaxy mass
models are built according to the most recent understanding of the
ETGs' structure, such as their stellar mass profile and their dark
matter content and distribution, as indicated by detailed modeling of
optical observations and by the main scaling laws (e.g., Cappellari et
al. 2006, Weijmans et al. 2009, Auger et al. 2010, Napolitano et
al. 2010, Shen \& Gebhardt 2010).  The aims are to address the
following questions: can the gas heating sources above account for the
observed $T$'s? how are the various input energy sources for the gas
used in the different flow phases? is there any relation between $T$
and the flow phase?

We present in Sect.~\ref{heat} the sources of mass and heating
for the hot ISM, in Sect.~\ref{esc} the conditions for the
gas to escape from the galaxy, in Sect.~\ref{mass} the galaxy mass
models, in Sect.~\ref{disc} the comparison between observed and
predicted temperatures, in Sect.~\ref{temp} the relation between gas
temperature and flow status, and in Sect.~\ref{concl} the conclusions.

\section{Sources of mass and heating for the hot gas}\label{heat}

\subsection{Gas Mass}\label{loss}

In ETGs the hot gas comes from stellar mass losses produced by evolved
stars, mainly during the red giant, asymptotic giant branch, and
planetary nebula phases, and by SNIa's, that are the
only ones observed in an old stellar population (e.g., Cappellaro et
al. 1999).  The first, more quiescent, type of losses originates ejecta that
initially have the velocity of the parent star, then individually
interact with the mass lost from other stars or with the hot ISM, and
mix with it (Mathews 1990, Parriott \& Bregman 2008).

For a galaxy of total stellar mass $M_*$, the evolution of the stellar
mass loss rate $\dot M_*(t)$ can be calculated using single burst
stellar population synthesis models (Maraston 2005), for a Salpeter
and for a Kroupa Initial Mass Function (IMF), assuming for example
solar abundance. So doing, at an age of 12 Gyrs, a rate is recovered
of $\dot M_* $= B $\times 10^{-11} \, L_B(L_{B,\odot})\,\,
M_{\odot}$yr$^{-1}$, where $L_B$ is the galactic B-band luminosity at an age of
12 Gyr, and B=1.8 or B=1.9 for the Salpeter or Kroupa IMF (see also Pellegrini 2011). This
value is in reasonable agreement with the average derived for nine
local ETGs from $ISO$ data (Athey et al. 2002) of $\dot M_*=7.8\times
10^{-12}\,\, L_B(L_{B,\odot})$ M$_{\odot}$yr$^{-1}$, an estimate based
on individual observed values that vary by a factor of $\sim 10$,
though, which was attributed to different ages and metallicities.

The total mass loss rate of a stellar population $\dot M$ is given by
the sum $\dot M=\dot M_*+ \dot M_{\rm SN}$, where $ \dot M_{\rm SN}$
is the rate of mass loss via SNIa events for the whole galaxy. $\dot
M_{\rm SN}$ is given by $\dot M_{\rm SN} =M_{SN}\, R_{\rm SN}$, where
$M_{SN}=1.4M_{\odot}$ is the ejected mass by one event, and $R_{\rm
SN}$ is the explosion rate. $R_{\rm SN}$ has been determined for
local ETGs to be $R_{\rm SN}=0.16 (H_0/70)^2 \times 10^{-12} \,L_B
(L_{B,\odot}) \, {\rm yr}^{-1} $, where $H_0$ is the Hubble constant
in units of km s$^{-1}$ Mpc$^{-1}$ (Cappellaro et al. 1999). More
recent measurements of the observed rates of supernovae in the local
Universe (Li et al. 2010) give a SNIa's rate in ETGs consistent with
that of Cappellaro et al. (1999). For $H_0=70$ km s$^{-1}$
Mpc$^{-1}$, one obtains $\dot M_{\rm SN}$ =2.2$ \times 10^{-13}
L_B(L_{B,\odot}) $ M$_{\odot}$ yr$^{-1}$, that is $\sim 80$ times
smaller than $\dot M_* $ derived above for an
age of 12 Gyr; therefore, the main source of mass for the hot gas is
provided by $\dot M_*$.  A reasonable assumption is that the gas is
shed by stars with a radial dependence that follows that of the
stellar distribution, so that the density profile of the injected
gas is $\rho_{gas}(r)\propto \rho_*(r)$, where $\rho_*(r)$ is the
stellar density profile.  This assumption is adopted hereafter, and
the characteristic temperatures presented below apply to a gas 
distribution following $\rho_{gas}(r)\propto \rho_*(r)$ (but see also Sect.~\ref{disc}).

\subsection{Heating from stellar motions and supernovae}\label{inj}

The material lost by stars is ejected at a velocity of few tens of km
s$^{-1}$ and at a temperature of $\lsim 10^4$ K (Parriott \& Bregman
2008), and is subsequently heated to high, X-ray emitting temperatures
by the thermalization of the stellar velocity dispersion, as it
collides with the mass lost from other stars, or with the ambient hot
gas, and is shocked. Another source of heating for the stellar mass
losses is provided by the thermalization of the kinetic energy of
SNIa's events. The internal energy given by these heating processes to
the unit mass of injected gas is $3kT_{inj}/2\mu m_p$ (with $k$ the
Boltzmann constant, $m_p$ the proton mass, $\mu m_p$ the mean particle
mass, with $\mu=0.62$ for solar abundance); $T_{inj}$ is
determined by the heating due to thermalization of the motions of the
gas-losing stars ($T_{star}$), and of the velocity of the SNIa's
ejecta ($T_{SN}$), and is written as (e.g., Gisler 1976, White \&
Chevalier 1983): 
\begin{equation} T_{inj}=T_{star}+T_{SN}={\dot M_*
T_{*}+\dot M_{SN} T_{ej} \over \dot  M } .  
\label{eq:tinj} 
\end{equation} 
Here $T_{*}$ is the equivalent temperature of the stellar motions (see
below), and $T_{ej}= 2\mu m_p E_{SN}/(3kM_{SN})$ is the equivalent
temperature of the kinetic energy $E_{\rm SN}$ of the SNIa's ejecta,
with $E_{\rm SN}= 10^{51}$ erg for one event (e.g., Larson 1974).
$T_{ej}$ can be calculated assuming that a factor $f$ of $E_{\rm SN}$
is turned into heat; $f<1$, since radiative energy losses from
expanding supernova remnants may be important, and values down to
$f=0.1$ have been adopted (Larson 1974, Chevalier 1974); a value of
$f=0.85$ could be not too far off for the hot diluted ISM of ETGs 
  (e.g., Tang \& Wang 2005). In this way, $T_{ej}= (f/0.85) 1.5\times 10^9$K.  From
approximating $\dot M \simeq \dot M_*$, and using the estimates of
Sect.~\ref{loss} for $\dot M_{SN}$ and $\dot M_*$ (for a Kroupa IMF)
at the present  epoch, one obtains
$T_{SN}\simeq 1.7(f/0.85)\times 10^7$ K.

The injection temperature $T_{inj}$ is then the sum of two
parts: one ($T_{SN}$) is independent of the position within the galaxy
where the gas is injected (e.g., independent of radius in spherical
symmetry), and is also constant from galaxy to galaxy (for fixed IMF
and age of the stellar population, and SNIa's rate); for each ETG it
can, though, evolve with time, if $\dot M_{SN}$ and $\dot M_*$ evolve
differently with time (Ciotti et al. 1991).  The other part
($T_{star}$) is instead basically independent of time, but has a
radial dependence, and changes with the galaxy structure, i.e., with
the total mass and its distribution. An average $T_*$
is obtained calculating the gas mass-weighted temperature gained
by the thermalization of the stellar random motions, $<T_*>$:
\begin{equation} 
<T_{*}>={1\over k} {\mu m_p \over M_*} \int 4 \pi
r^2 \rho_*(r) \sigma^2 (r) \,dr,
\label{eq:tvir} 
\end{equation} 
where $\sigma (r)$ is the
one-dimensional velocity dispersion of the stars.  The integral term
in Eq.~\ref{eq:tvir} is the same that gives the kinetic energy
associated with the stellar random motions [$E_{kin}=1.5 \int 4 \pi
r^2 \rho_*(r) \sigma^2 (r) \,dr $], and that enters the virial theorem
for the stellar component; the mass-weighted temperature in
Eq.~\ref{eq:tvir} is then often called ``gas virial temperature''.
For a galaxy mass model made of stars and dark matter,
characterized by $ {\cal R}=M_h/M_*$, where $M_h$ is the total dark
mass, and $\beta=r_h/r_*$, with $r_h$ and $r_*$ the scale radii of the
two mass distributions, $<T_*>$ can be
expressed using the central velocity dispersion $\sigma_c$ as
$<T_*>=\mu m_p\, \sigma_c^2 \Omega ({\cal R},\beta)/k$ (e.g., Ciotti \& Pellegrini 
1992). The
function $\Omega $ increases mildly for larger $\cal{R}$ and for lower
$\beta$, that is for a larger amount of gravitating mass or a higher
mass concentration, but always $\Omega <1$, since
$\sigma (r) $ has in general a negative radial gradient (e.g., Sect.~\ref{mass}
and Fig.~\ref{f1} below).  $<T_*>$ is then proportional to
$\sigma_c^2$, and a
simplified version of the
virial temperature in Eq.~\ref{eq:tvir} that is often used is 
$T_{\sigma}=\mu m_p\sigma_c^2/k$; $T_{\sigma}$ of course overestimates
the true $<T_*>$.

The mass-averaged injection temperature is finally given by
\begin{equation} <T_{inj}> =
<T_{*}> + 1.7(f/0.85)\times 10^7 \,\, K,  
\label{eq:tav}
\end{equation} 
where in general the second term dominates, as is shown in Sect.~\ref{disc} below.

\subsection{Heating during Infall}\label{egp}

In case of mass losses flowing to the galactic center, the gas can be 
heated due to infall in the galactic potential and adiabatic
compression; this process is sometimes referred to as ``gravitational
heating''.  The average change in gravitational energy per unit gas mass
inflowing through the galactic potential down to the galactic center is
\begin{equation}
E_{\rm grav}^+ ={1\over M_*}\int ^{\infty}_{0}4\pi r^{2}\rho_*(r)
[\phi (r) - \phi(0)] dr,
\label{eq:lgp}
\end{equation}
for galaxy mass distributions with a finite value of $\phi(0)$ (see also 
Ciotti et al. 1991).  One
can define a temperature equivalent to the energy in Eq.~\ref{eq:lgp}
as $<T_{\rm grav}^{+}>=2\mu m_p\,E_{\rm grav}^{+}/3k$.  As
$<T_*>$, also $<T_{\rm grav}^+> $ is $\propto \sigma_c^2$, and
increases for larger ${\cal R}$ and smaller $\beta$, which, for
inflowing gas, can be understood as a larger gas heating by
compression during infall for a larger dark matter amount or its
higher concentration.

Not all of $E_{\rm grav}^+$ can be available for heating, though.  If
the inflow keeps quasi-hydrostatic, then, by the virial
theorem, the energy radiated away is roughly one-half of the change in
the gravitational potential energy,
and that available for the heating of the gas is the remaining
half\footnote{The energy lost in radiation and that
  converted into heat are actually each equal to $0.5 E_{\rm grav}^+ $ for a
  self-gravitating gas; for gas in an external potential, the result
  should remain roughly valid (Binney \& Tremaine 1987).}
 (i.e., $\sim 0.5 E_{\rm grav}^+$). Actually, the energy
  available for heating will be much less than this. Inflows are caused by the
  radiative losses produced by the accumulation of the
  stellar mass return, that makes the cooling time lower than the
  galactic age; in the central regions, within a radius of $\sim 1$ kpc, 
the cooling time can be as short as $\lsim 10^8$ yr, even
shorter than  the infall time (e.g., Sarazin \& White 1988, Pellegrini
2011). In these conditions, the gas departs from a slow
inflow, becomes very dense and
supersonic close to the center, and cools rapidly down to low
temperatures, so that $> 0.5 E_{\rm grav}^+$ is radiated away or goes
into kinetic energy of condensations (Sarazin \& Ashe 1989).  Furthermore,
there is the possibility that not all the gas
reaches hot the galactic central region, if thermal instabilities develop
and produce drop-outs from the flow; if gas cools and condenses out of
the flow at large radii, then $E_{\rm grav}^+$ can be much lower than
in the definition above, and heating due to infall in the
gravitational potential is ``lost'' (Sarazin \& Ashe 1989).  In
conclusion, without a precise knowledge of how to compute $E_{\rm
  grav}^+$ (which depends on the radius at which the injected gas
drops below X-ray emitting temperatures), and about what fraction of $E_{\rm
  grav}^+$ is radiated or goes into kinetic energy of the condensations,
$<T_{\rm grav}^{+}>$ remains a reference value; a more direct use
can instead be made of the analogous temperature for escape $<T_{\rm
  grav}^{-}>$ introduced in Sect.~\ref{esc} below.

\subsection{Heating from a central MBH}\label{mbh}

Due to the presence of a central MBH in ETGs, another potential source
of heating for the gas could be provided by nuclear accretion.  This
subject has been studied intensely recently, through both observations
and modeling, and it appears that the energy provided by accretion is
of the order of that needed to offset cyclically the cooling of
the inflowing gas in the central regions of gas-rich ETGs (e.g.,
B\^irzan et al. 2004, Million et al. 2010).  Therefore, the central
MBH is believed to be a heat source that balances the radiative losses
of the gas, acting mostly in the central cooling region.  In gas poor
ETGs the nuclear accretion energy is far lower due to the very small
mass accretion rate, if present (Pellegrini et al. 2007), and the
absorption of the energy output from accretion is likely not
efficient.  Given the role of the MBH outlined above, possible energy
input from the MBH will not be considered as a source of global
heating for the gas.

\section{Conditions for Escape}\label{esc}

Another characteristic temperature for comparison with observed $T$
values is the temperature with which the gas can escape from the
galaxy.  Assuming that the flow is stationary and
adiabatic, the Bernoulli constant on each streamline along which the
gas flows out of the galaxy must be positive.  The Bernoulli
equation with the minimum energy for escape is written as
$H(r)+v^2(r)/2+\phi(r)=0$, where $H={\gamma\over \gamma -1}{kT\over
  \mu m_p}= {c_s^2\over \gamma - 1}$ is the enthalpy per unit gas
mass, $\gamma$ is the ratio of specific heats, $c_s$ is the sound velocity, and $v$ is the flow velocity.
Integrating over the galaxy volume and gas-mass averaging,
this condition becomes:
\begin{equation}
\int ^{\infty}_{0} 4\pi r^{2}\rho_*(r)H(r) dr + {1\over 2}
\int ^{\infty}_{0} 4\pi r^{2}\rho_*(r) v^2(r) dr 
=M_* E_{\rm grav}^-,
\label{eq:bern}  
\end{equation}
where the escape energy
\begin{equation} 
E_{\rm grav}^-=-{1\over M_*} \int ^{\infty}_{0}4\pi r^{2}\rho_*(r)\phi(r) dr
\label{eq:lm} 
\end{equation}
is the average energy required to remove from the galaxy the unit gas mass.
$E_{\rm grav}^{-}$ gives a minimum energy requirement, since energy
losses due to cooling may be present; but these are not
important for outflows that typically have a low density (e.g.,
Sect.~\ref{sna} below).
The escape temperature equivalent to $E_{\rm grav}^-$ 
is $<T_{\rm grav}^{-}>=2\mu m_p\,E_{\rm grav}^{-}/3k$. The condition
for the minimum energy for escape can then be
translated into a condition for the injection temperature of the gas
$<T_{inj}>$ to be larger than $<T_{\rm grav}^{-}>$.
In the simple case that
$\phi(r)$ is due only to one (stellar) mass component,
from $<T_*>=2\mu m_p E_{kin}/3kM_*$ and the virial theorem
(Sect.~\ref{heat}), one derives that $<T_{\rm grav}^{-}>=4<T_*>$.
The estimates of $E_{\rm grav}^{-}$ in Eq.~\ref{eq:lm}  and then of
$<T_{\rm grav}^{-}>$ will be calculated below for a general mass model (e.g.,
made by the superposition of stars and dark matter, with a different
radial distribution). 
In previous works, the sufficient condition for
the existence of a galactic wind 
was that the injection temperature exceeded an ``escape
temperature'',  defined as $2T_{\sigma}$ (White \&
Chevalier 1983), or "twice the equivalent dark halo temperature"
(Loewenstein \& Mathews 1987), coupled with the request for the
radiative cooling time in the central part of the galaxy to
be longer than the time required to flow out of this
region. These conditions are similar to imposing that $<T_{inj}>$ 
exceeds $<T_{\rm  grav}^{-}>$ as derived above.

In principle the gas can escape with different combinations of $v$ and
$T$, and the observed $T$ should be close\footnote{For example, we
recall two approximations made here with respect to the case of real ETGs: 
the total gas profile may be different from that of the stars, and 
the flow has a time-continuous distributed mass and energy input. The
first of these points will be further discussed in Sect.~\ref{disc}.}
 to that entering $H$ in Eq.~\ref{eq:bern}.
There are two extreme cases for the value of the flow velocity $v$
with respect to $c_s$ (i.e., to the temperature).  One is when the material is brought
to infinity keeping a subsonic velocity, then the minimum energy
requirement becomes $H\approx -\phi$; neglecting the kinetic
term in Eq.~\ref{eq:bern}, one then obtains a characteristic gas-mass
averaged subsonic escape temperature:
\begin{equation} 
<T_{\rm esc}^{sub}>={2\mu m_p\over 5k M_*} \int ^{\infty}_{0} 4\pi
r^{2}\rho_*(r)\phi(r) dr={3\over 5} <T_{\rm grav}^{-}>.  
\label{eq:tent} 
\end{equation} 
In a general case, $H\approx -\phi$ gives for $T$ a larger requirement
than obtained when $v$ is not neglected; also the partition of the gas
energy between enthalpy and kinetic energy can vary with radius within
a galaxy. All this means
that $<T_{\rm esc}^{sub}>$ represents a fiducial upper limit to the
observed temperatures of outflowing gas: if the kinetic energy of the
flow is important, then the actual gas temperature will be lower (the
stronger the outflow, with respect to $c_s$, the cooler the gas). At
the opposite extreme case where the ``temperature'' contribution to
the gas energy is minor and that of the velocity is dominant, the
Bernoulli equation reduces to $v_{esc}^2/2+\phi=0$; this gives the
usual escape velocity of a unit mass from a potential well: $v_{esc}
(r)=\sqrt{ 2 |\phi(r)|}$.

Finally,  $<T_{\rm grav}^{-}>$ and $<T_{\rm esc}^{sub}>$ have the same
dependence as $<T_*>$ on $\sigma_c^2, {\cal R},\beta$.  For the
representative galaxy mass models used here (Sect.~\ref{mass},
Fig.~\ref{f2}), $E_{\rm grav}^+ =(1.7 - 2.3) E_{\rm grav}^-$; since
$\lsim 0.5 E_{\rm grav}^+$ can be converted into heat, then the
corresponding temperature gained by infall will be $T_{infl}\lsim
<T_{\rm grav}^{-}>$.

\section{Galaxy mass models}\label{mass}

In this work the exact values of $<T_*>$, $<T_{\rm esc}^{sub}>$ and $<T_{\rm
  grav}^{-}>$ are calculated as a function of $\sigma_c$, for a
series of representative 3-component galaxy mass models, made by the
superposition of a stellar distribution and a dark matter halo, to
which a central MBH is added. The stellar density
profile is given by the deprojection of a S\'ersic law with index
$n=4$ or 5, as appropriate for ETGs of the luminosities considered in
this work (e.g., Kormendy et al. 2009). The mass of the MBH is 
$M_{BH}=10^{-3}M_*$, in agreement with the Magorrian et al. (1998) relation. The dark halo has a
Navarro et al. (1997; NFW) profile [$\rho_h\propto 1/(
r/r_h)(1+r/r_h)^2$, with $r_h$ the scale radius, and
truncated at large radii], and a total mass
$M_h$. For each $\sigma_c$, the defining parameters of the
stellar mass model were chosen accordingly to the observational
constraints that  $L_B$ follows the
Faber-Jackson relation, and that $L_B$, $\sigma_c$ and the effective
radius $R_e$ lie on the Fundamental Plane of ETGs (e.g., Bernardi et
al. 2003).  The free parameters defining the dark matter were chosen
in agreement with the results from dynamical modeling of the observed motions
of stars, planetary nebulae and globular clusters at small and large
radii; these indicate that the dark matter begins to be dynamically
important at 2--3$R_e$ (e.g., Saglia et al. 1992, Cappellari et al. 2006, 
Weijmans et al. 2009, Shen \& Gebhardt 2010). This requires that
$\beta =r_h/R_e>1$, and ${\cal R}=M_h/M_*=3$ or 5 (the latter value
corresponding to the baryon-to-total mass ratio of WMAP,
Komatsu et al. 2009).  By solving numerically the Jeans equations for
the three mass components in the isotropic orbits case (e.g., Binney
\& Tremaine 1987), these choices produce $M_*$, $M_h$ and ${\cal R}_e$
(the
dark-to-luminous mass ratio within $R_e$).  Reasonable
values of $M_*/L_B=(4-10) \, M_{\odot}/L_{B,\odot}$, and ${\cal R}_e =
0.2-1.0$ are obtained. The main properties of a few representative
mass models are shown in Fig.~\ref{f1}.

For a consistent comparison between observed $T$'s and the
characteristic temperatures derived for the mass models, the
central stellar velocity dispersion $\sigma_c$ must be the same
for observed ETGs and models.  Typically, for nearby well observed
ETGs, the value of $\sigma_c$ is that of the projected and
luminosity-weighted average within an aperture of radius $R_e/8$.
Therefore, when defining a mass model, the chosen value of $\sigma_c$
was assigned to this quantity.  Finally, streaming motions as stellar
rotation are not considered in these models (possible heating from 
these motions is discussed by Ciotti \& Pellegrini 1996).

\section{Discussion}\label{disc}

We investigate here the relationship between the observed $T$'s and
those expected from the various sources of heating (stellar motions,
gravitational potential, SNIa's), or during the escape of the hot gas.
For this purpose, Fig.~\ref{f2} shows the run with $\sigma_c$ of the
various temperatures defined in Sects.~\ref{inj}, \ref{egp},
and~\ref{esc}, together with the distribution of the observed $T$
values from BKF (Sect.~\ref{intro}); for the BKF sample, the value of
$\sigma_c$ is the luminosity weighted average within $R_e/8$, and has
been taken from SAURON studies for 12 ETGs (Kuntschner et al.  2010),
for the remaining cases from the references in the
Hyperleda catalog (see Tab.~\ref{tab1}).

The temperatures defined in Sects.~\ref{inj}, \ref{egp},
  and~\ref{esc} are mass-weighted averages, which is required when
  discussing energetic aspects of the gas (e.g., the energy required
  for escape as measured by $<T_{\rm grav}^->$ compared with the input
  energy from SNIa's).  When a direct comparison is made with observed
  $T$'s, it must be noted that the latter coincide with mass-weighted
  averages only if the ISM has everywhere one single temperature
  value; if the gas is multi-phase, or its temperature profile has a
  gradient, a single $T$ value measured from the spectrum of the
  integrated emission will be close to an emission-weighted average
  (e.g., Ciotti \& Pellegrini 2008, Kim 2011).  This means that, since
  the densest region is the central one, the measured $T$'s tend to be
  closer to the central values than the mass-weighted ones.  The
  temperature profiles observed with $Chandra$ change continuously in
  shape, as the emission-weighted average $T$ decreases from $\lsim 1$
  keV to $\sim 0.3 $ keV: they switch from a flat central profile that
  increases outward of $\sim 0.5 R_e$, to a quasi-isothermal profile,
  to a profile with a negative gradient (Diehl \& Statler 2008, Nagino
  \& Matsushita 2009).  Therefore, the lowest observed $T$'s,
  presumably associated with the last category, may be larger than
  mass-weighted values, intermediate $T$'s may be the closest to
  mass-weighted averages, while the largest observed $T$'s may be
  lower than mass-weighted averages. Another aspect to recall is that
  the temperatures defined in Sects.~\ref{inj}, \ref{egp},
  and~\ref{esc} refer to a gas distribution with $\rho_{gas} \propto
  \rho_*$; this is appropriate for the continuously injected gas
  (e.g., for $<T_{inj}>$), while it may be less accurate when
  comparing observed $T$'s with $<T_{\rm esc}^{sub}>$, or when
  discussing the energetics of the whole gas content of an ETGs by
  means of $<T_{\rm grav}^+>$ and $<T_{\rm grav}^->$, since the bulk
  of the hot ISM may have a different distribution from the stars.
  For example, the observed X-ray brightness profile of gas-rich ETGs
  was found to follow the optical one, which was taken as evidence
  that roughly $\rho_{gas}\propto {\sqrt \rho_*}$ (e.g., Sarazin \&
  White 1988, Fabbiano 1989).  For gas-poor ETGs hosting galactic
  winds, the modeling shows that the profile $\rho_{gas} (r)$ will
  again be shallower than $\rho_* (r)$, though not as much as in the
  previous case (see, e.g., White \& Chevalier 1983).  If $\rho_{gas}$
  has a flatter radial profile than $\rho_*$, then it is easy to show
  that its mass-weighted $<T_{\rm grav}^+>$ will be larger than
  derived using Eq.~\ref{eq:lgp}, and its mass-weighted $<T_{\rm
    grav}^->$ and $<T_{\rm esc}^{sub}>$ will be lower than derived
  using Eqs.~\ref{eq:lm} and~\ref{eq:tent}.  In conclusion, the
  comparison of observed $T$'s with mass-weighted expectations is the
  best that can be done currently, in a general analysis as that of
  the present work, though with the warnings above. Note, however,
  that all arguments and conclusions below remain valid or are
  strenghtened, when taking into account the above considerations
  about observed $T$'s, or about the modifications to $<T_{\rm
    grav}^+>$, $<T_{\rm grav}^->$ and $<T_{\rm esc}^{sub}>$.

\subsection{Observed and predicted temperatures in the $T-\sigma_c$ plane}\label{obs}

\begin{figure*}
\hskip -0.4truecm
\includegraphics[height=0.43\textheight,width=0.54\textwidth]{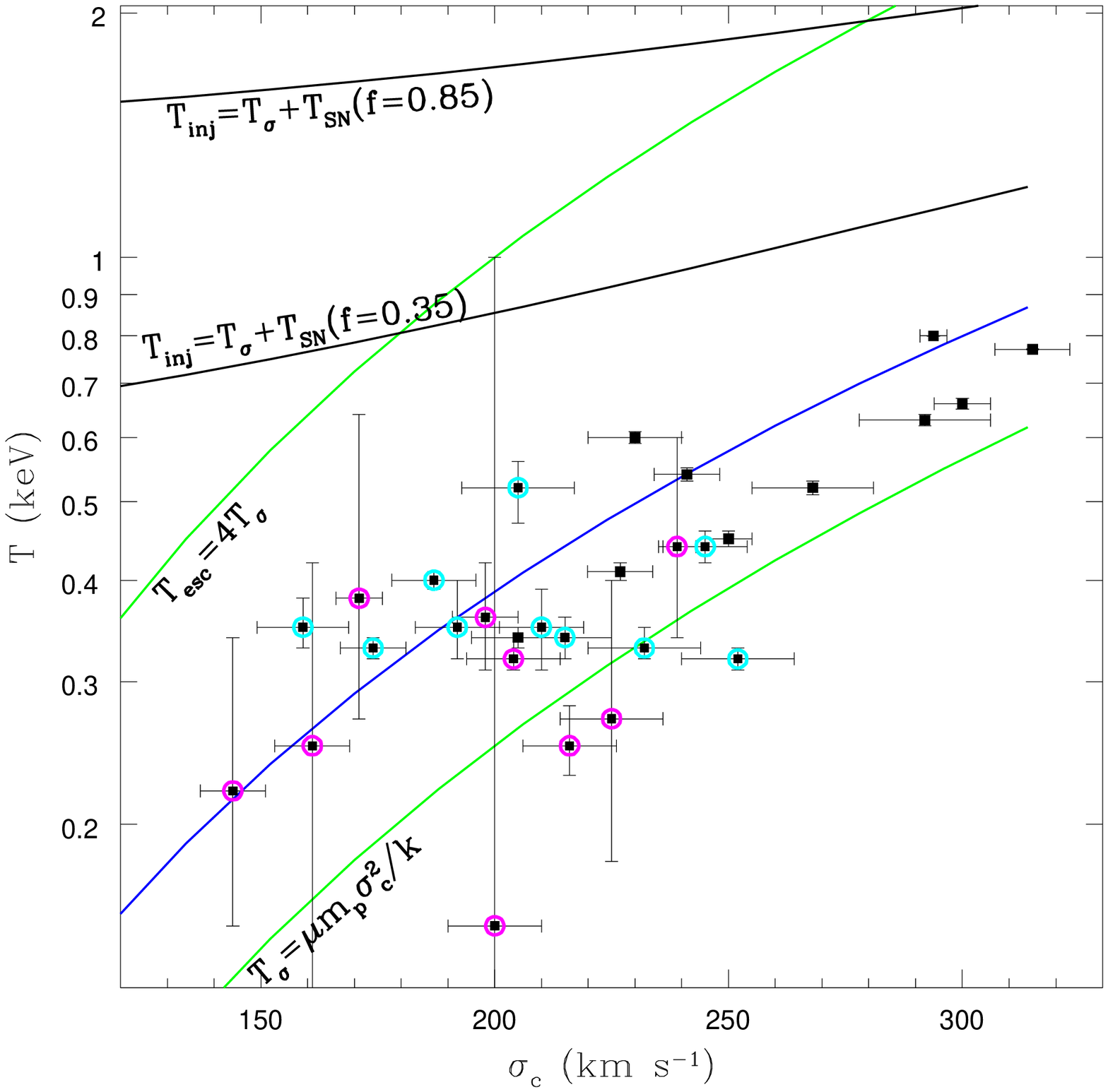}
\hskip -0.2truecm
\includegraphics[height=0.43\textheight,width=0.54\textwidth]{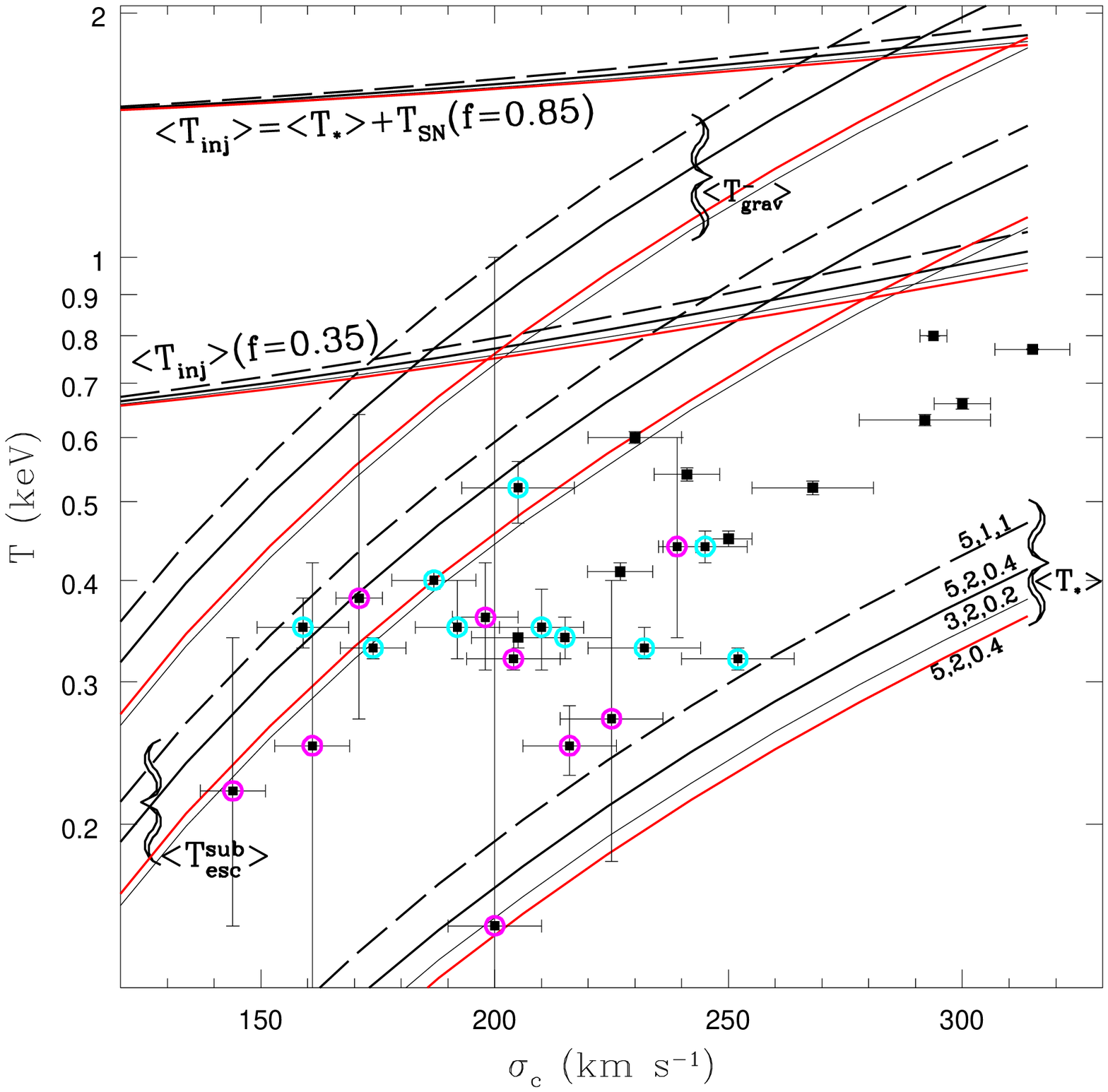}
\caption{The relationship between the observed gas temperature (from
BKF) and $\sigma_c$ (see Sect.~\ref{obs}). Symbols surrounded by a
magenta and cyan circle have respectively $10^{38}$ erg s$^{-1}<L_X<1.5\times 10^{39}$ erg
s$^{-1}$, and $1.5\times 10^{39}$ erg s$^{-1}<L_X<1.2\times 10^{40}$
erg s$^{-1}$; all other ETGs have larger $L_X$. Left panel: in green
$T_{\sigma}$ (Sect.~\ref{heat}), and the simple estimate of
$4T_{\sigma}$ for the escape temperature (Sect.~\ref{esc}); in blue
the best fit $\sigma_c\propto T^{0.56\pm 0.09}$ found from $ROSAT$
data (O'Sullivan et al. 2003); in black two cases of $T_{inj}$
(Eq.~\ref{eq:tav}), calculated using $T_{\sigma}$. Right panel: 1)
$<T_*>$ (Eq.~\ref{eq:tvir}, lowest bundle of lines), calculated for
four representative galaxy mass models (made of MBH$+$stars$+$dark
halo), with a S\'ersic index $n=4$ (black lines) or $n=5$ (red line),
and the dark matter parameters ${\cal R},\,\beta,\,{\cal R}_e$
indicated on each curve (Sect.~\ref{mass}); 2) $<T_{\rm grav}^->$ and
$<T_{\rm esc}^{sub}>$ (Sect.~\ref{esc}) for the same mass models
adopted for $<T_*>$, with the corresponding line type and color; 3)
$<T_{inj}>$ calculated using $<T_*>$, with the corresponding line type
and color, and $f=0.85$ or $f=0.35$ in Eq.~\ref{eq:tav}.}
\label{f2}
\end{figure*}

In the left panel of Fig.~\ref{f2} the observed $T$'s are 
compared with approximate estimates of the stellar temperature
$T_{\sigma}$, of $T_{inj}$, and of the escape temperature $4T_{\sigma}$
(Sect.~\ref{esc}).  The gas luminosity is also indicated with
different colors, having grouped the $L_X$ values in three ranges,
chosen to have a roughly equal number of ETGs in each range. This
grouping gives an indication of the gas flow status, based on previous
works: a galactic wind leaving the galaxy with a supersonic velocity
has $L_X<10^{38}$ erg s$^{-1}$ (e.g., Mathews \& Baker 1971,
Trinchieri et al. 2008), global subsonic outflows and partial winds
can reach $L_X\sim 10^{40}$ erg s$^{-1}$ (Ciotti et al. 1991,
Pellegrini \& Ciotti 1998), and a central inflow becomes increasingly
more important in ETGs of increasingly larger $L_X$. Magenta ETGs
($10^{38}$ erg s$^{-1}<L_X<1.5\times 10^{39}$ erg s$^{-1}$) should
then host winds, subsonic outflows, and partial winds with a very
small inflowing region of radius $<100$ pc; cyan ETGs ($1.5\times
10^{39}$ erg s$^{-1}<L_X<1.2\times 10^{40}$ erg s$^{-1}$) should host
subsonic outflows and partial winds with an increasingly larger
inflowing region (of radius up to a few hundreds pc); black ETGs are
hot gas-rich and mostly inflowing \footnote{ In the X-ray faintest
    ETGs, the gas emission $L_X$ is $\sim (1-2)$ times the integrated
    emission from the population of weak unresolved stellar sources
    (Sect.~\ref{intro}), generally referred to as AB$+$CVs.  BKF
    derived an AB$+$CV emission model by jointly fitting the M31 and
    M32 spectra, and then tested how the measurement of the gas
    properties may be affected by the adopted AB$+$CV model (their
    Sect. 3.1).  By fitting with different AB$+$CV models, all within
    the uncertainties in the adopted one, the measured $T$ changed
    only negligibly, for the 6 lowest-$L_X$ ETGs of their sample; a
    systematic uncertainty of 10\% -- 20\% was found for the gas flux
    (much lower for the X-ray brighter ETGs). }. 
The
best fit found for X-ray bright ETGs is also shown in the left panel
(O'Sullivan et al. 2003), and gives a good representation of the
distribution of observed $T$'s down to a range of low temperatures and
gas contents never explored before. The slope of the fit ($T\propto
\sigma_c^{1.79}$) and that of the $T_{\sigma }\propto \sigma_c^2$
relation are similar, with the fit being shallower; this could be
due to the fact that not all heating sources depend on $\sigma_c^2$,
see, e.g., the important SNIa's contribution in $T_{inj}$, that
produces a much flatter run of $T_{inj}$ with $\sigma_c$
(Fig.~\ref{f2}). The fit was mostly based on gas-rich ETGs, whose
$T$'s show a trend with $\sigma_c$ closer to that of $T_{\sigma}$
(an aspect further addressed below in Sects.~\ref{gravhea}
and~\ref{temp}), while gas-poor ETGs depart most from it, since their
$T$'s change little for largely varying $\sigma_c $ (as found by BKF;
see Sect.~\ref{reduc} below).

The right panel of
Fig.~\ref{f2} shows the 
stellar temperature $<T_*>$, the injection temperature $<T_{inj}>$,
the escape temperature $<T_{\rm grav}^->$, and the characteristic
temperature for slowly outflowing gas $<T_{\rm esc}^{sub}>$,
calculated for a set of representative galaxy mass models (Sect.~\ref{mass}).
At any fixed $\sigma _c$ and S\'ersic index $n$,
$<T_*>$, $<T_{\rm grav}^->$ and $<T_{\rm esc}^{sub}>$ are larger for
larger galaxy mass (${\cal R}$), and mass concentration (smaller
$\beta$)\footnote{All the rest equal, these temperatures are also
  larger for smaller $n$, due to the galaxy being more massive to
  reproduce the same $\sigma_c$, since the stellar mass profile is
  less steep (e.g., Fig.~\ref{f1}).}.  The dashed lines represent a
reasonable upper limit to the values of each of the characteristic
temperatures, since they correspond to the most massive model ETGs,
with the most concentrated dark matter allowed for by recent studies
(Sect.~\ref{mass}). The $<T_{\rm grav}^->$ curves lie below
the simple approximation of the escape temperature
given by $4T_{\sigma}$;  $<T_{\rm  grav}^->=4.8<T_*>$, 
for ${\cal R}=3$, and $\simeq 5.2<T_*>$, for the three cases with ${\cal R}=5$.

As expected, all the $<T_*>$ curves lie below $T_{\sigma}$, that
overestimates the kinetic energy associated with the stellar random
motions (Sect.~\ref{inj}).  Note that, from the virial theorem,
$<T_*>$ is independent of orbital anisotropy, that just redistributes
differently the stellar heating within a galaxy; the presence of
ordered rotation in the stellar motions, instead, requires a more
careful consideration. For any fixed galaxy mass model, this rotation
would leave the total stellar heating unchanged or lower it, depending
on whether the whole stellar streaming motion is converted into heat,
or just a fraction of it (Ciotti \& Pellegrini 1996). For the worst
case that the stellar rotational motion is not thermalized at all, and
the galaxy is a flat isotropic rotator, $<T_*>$ in Fig.~\ref{f2}
should be an
overestimate of $\sim 30$\% of the temperature corresponding to the
stellar heating (Ciotti \& Pellegrini 1996); the possible reduction of
$<T_*>$ to be considered should be lower than this, as far as 
the massive ETGs in Fig.~\ref{f2} are less flattened
and more pressure supported systems (e.g., Emsellem et al. 2011).

All observed $T$'s are located above $<T_*>$; thus, additional heating with respect to
the thermalization of the stellar kinetic energy is needed, as noticed
previously using $T_{\sigma}$ (e.g., Davis \& White 1996, BKF).  The
gas could retain memory of its injection temperature, and have the
additional infall heating, as examined in Sects.~\ref{gravhea}
and~\ref{sna} below.

Finally, the values of $<T_{inj}>$ for $f=0.85$ are by far the largest
temperatures of Fig.~\ref{f2}, larger than $<T_{\rm grav}^->$ up to 
$\sigma_c\sim 250$ km s$^{-1}$; therefore, SNIa's should cause the escape 
of the gas for all ETGs up to this  $\sigma_c$, since the gas at every 
time is injected with an energy larger than required to leave the 
galaxy potential. This expectation is fulfilled by all ETGs with $\sigma_c 
\lsim 200$ km s$^{-1}$: their X-ray properties (a low $L_X$, and $T$'s of 
the order of $<T_{\rm esc}^{sub}>$) agree well with what expected if 
outflows are important in them. This result had been
  suggested previously based on the low observed $L_X$; now for the
  first time it can be confirmed based on the observed $T$ values.  At
  $\sigma_c > 200$ km s$^{-1}$, instead, ETGs may have $L_X$ far
  larger than expected for outflows (black symbols), and most 
ETGs where likely outflows are important (magenta or cyan symbols) 
have $T$ much lower than $<T_{\rm esc}^{sub}>$; these findings are discussed in
  Sect.~\ref{reduc} below.

\subsection{ Gravitational heating in
  gas-rich ETGs}\label{gravhea}

We examine here the possibility that the additional heating with
respect to the thermalization of the stellar kinetic energy is
provided by infall heating and SNIa's.  Davis \& White (1996) assumed
that in all ETGs the hot gas is inflowing, and suggested that the
observed temperatures are larger than $T_{\sigma}$ just due to the
luminous parts of ETGs being embedded in dark matter halos dynamically
hotter than the stars; i.e., a form of ``gravitational potential'' way
for the gas heating was invoked. This way can consist of an effect of
the dark halo on the stellar motions, that are then thermalized, or
directly on the gas during infall (e.g., via $E_{\rm grav}^+$). 
The first possibility is excluded by the $<T_*>$ curves in
Fig.~\ref{f2}, that are always lower than $T_{\sigma}$, and that,
through the Jeans equations, include the effect of a massive dark halo
consistent with the current knowledge of the ETGs' structure.  In the
second possibility of heating from gas infall, $E_{\rm grav}^+$ is
indeed potentially an important source of heating, that increases with
the amount and concentration of the dark matter. This can be judged
from Fig.~\ref{f2}, after considering that $<T_{\rm grav}^->\sim
5<T_*>$, and that the temperature possibly attainable from infall was
estimated to be $T_{infl}\lsim <T_{\rm grav}^{-}>$ (end of
Sect.~\ref{esc}). Note that $T_{infl}$ (if it behaves as $<T_{\rm
  grav}^{+}>$) could be $\propto \sigma_c^2$, a trend close to that
shown by the $T$'s of gas-rich ETGs (BKF; see also Sect.~\ref{temp}
below).

Inflowing ETGs can also benefit of the SNIa's energy input; for the
unit mass of injected gas, this is written as
$E_{SN}^{tot}=R_{SN}E_{SN}/ \dot M_*$. Both $E_{\rm grav}^+$ and
$E_{SN}^{tot}$ then contribute to the required additional thermal
energy with respect to that gained from the stellar random motions,
i.e., to $\Delta E_{th}=3k(T-<T_*>)/2\mu m_p$.  $E_{\rm grav}^+$ and
$E_{SN}^{tot}$ can be in large part radiated in gas-rich ETGs, but
they seem to far exceed the required $\Delta E_{th}$.  For example,
for the highest-$L_X$ of Fig.~\ref{f2}, $\Delta E_{th}\sim (1-2)\times
10^{48}$ erg $M_{\odot}^{-1}$ (i.e., $\sim$0.2--0.4 keV), when adopting an average galaxy mass model as that of
the thick black line in Fig.~\ref{f2}. The energy spent in radiation
can be estimated, in a stationary situation, as $L_X/\dot M_*$ (per
unit injected gas mass); using $L_X$ from BKF and deriving $\dot M_*$
as in Sect.~\ref{loss}, for the same distances in BKF and galactic
B-magnitudes given by Hyperleda, for the gas-rich ETGs one finds that
$L_X/\dot M_*$ ranges between $(0.5-3.4)\times 10^{48}$ erg
$M_{\odot}^{-1}$. The energy available is far larger than the sum of
$\Delta E_{th}$ and $L_X/\dot M_*$: $E_{SN}^{tot}=7.3\times
10^{48}(f/0.85)$ erg $M_{\odot}^{-1}$, and $0.5E_{\rm grav}^+$ ranges
from $4\times 10^{48}$ erg $M_{\odot}^{-1}$ ($\sigma_c\sim 220 $ km
s$^{-1}$) to $8\times 10^{48}$ erg $M_{\odot}^{-1}$ ($\sigma_c\sim 300
$ km s$^{-1}$), for the mass model with the thick black line in
Fig.~\ref{f2}. These results are detailed in Fig.~\ref{f3}, where the
values of $\Delta E_{th}$ and $L_X/\dot M_*$ for each galaxy are
shown, together with various combinations of $E_{\rm grav}^+$ and
$E_{SN}^{tot}$. In conclusion, additional input energy for the gas to
account for the observed $T$'s of gas-rich ETGs seems available in a
sufficient amount,  even if $f$ were to be $<0.85$.

\subsection{Outflows and SNIa's heating }\label{sna}

The gas is not mostly inflowing in all ETGs, while it is hotter than
$<T_*>$ in all of them. When in outflow, the radiative losses are far
smaller, but energy is spent in extracting the gas from the galaxy and
giving it a bulk velocity.  We discuss here the possibility of heating
from the SNIa's energy input to account for the observed $\Delta
E_{th}$ of ETGs where outflows are likely important (those with low/medium
$L_X$, magenta and cyan symbols in Fig.~\ref{f2}).

We assume that the SNIa's energy is used for the uplift of the gas and
the kinetic energy with which it escapes from the galaxy, and,
neglecting radiative losses, that all the remaining part is available
to account for the observed $\Delta E_{th}$.  Then, the energy balance
per unit mass of injected gas is $E_{SN}^{tot}= E_{\rm grav}^- +
E_{out} +\Delta E_{th} $, where $\Delta E_{th}$ is the same as in
Sect.~\ref{gravhea}, and $E_{out}=v^2_{out}/2$ is the mass-averaged
kinetic energy of the escaping material per unit gas mass.
Figure~\ref{f4} shows $\Delta E_{th}$ derived from this balance, for
$v_{out}=c_s$, and $c_s$ calculated for $\gamma=5/3$ and $kT=0.3
$ keV, a temperature of the order of that observed for ETGs likely in
outflow (Fig.~\ref{f2}). Adopting $v_{out}\sim c_s$ (independent of $\sigma_c$)
produces an $E_{out}$ on the upper end of those expected\footnote{For example, in
a wind solution, the terminal (i.e., the largest) velocity of the flow
is roughly the central sound speed (White \& Chevalier 1983); moreover, in
this solution the gas is likely to be already too ``fast'' with respect to that of most 
magenta ETGs in Fig.~\ref{f2}, due to their $L_X$ (e.g., Trinchieri et al. 2008).}, and
then the estimate of $\Delta E_{th}$ may be biased low. With
$v_{out}\sim c_s$, $E_{out}$ is just $\sim 0.1 (f/0.85)
E_{SN}^{tot}$. The energy needed for
 gas extraction $E_{\rm grav}^-$, for the same  galaxy mass model used in Sect.~\ref{gravhea},
varies instead from $\sim 1/3(f/0.85)E_{SN}^{tot}$ for $\sigma_c = 150$
km s$^{-1}$, to $\sim E_{SN}^{tot}$  for $\sigma_c = 250$
km s$^{-1}$; this explains the strong dependence of the
predicted $\Delta E_{th}$ on $\sigma_c$ in Fig.~\ref{f4}.  It is clear
from this figure that for $f=0.85$ SNIa's can account for the needed heating in all
ETGs with low/medium
$L_X$; for $f=0.35$, instead, the temperature increase
would fall short of what required for all ETGs. 
In Fig.~\ref{f4}, the energy losses due to radiation are also shown; their 
small size supports the hypothesis that in most cases they do not affect significantly
the energy budget of the gas.

Given the flat distribution of the observed points in Fig.~\ref{f4},
and the steep behavior of the curves predicting $\Delta E_{th}$, the
value of $f$ required to account for the observed $\Delta E_{th}$ increases
with $\sigma_c$.  In particular, the value of $f\sim 0.85$
that is  required at high $\sigma_c$ would produce an expected
$\Delta E_{th}$ at low $\sigma_c$ that is larger than observed.
 A possible solution could reside
in the efficiency of the SNIa's energy mixing process. In
  massive, gas-rich ETGs, SNIa's bubbles should disrupt and share
  their energy with the local gas within $\sim 3\times 10^6$ yr
  (Mathews 1990); for a Milky Way-size bulge in a global wind,
  instead, 3D
  hydrodynamical simulations of discrete heating from SNIa's suggest
  a non-uniform thermalization of the SNIa's energy, with
  overheated gas by a SNIa explosion at the bulge center that is
  advected outwards, carrying a large fraction of the SNIa energy with
  it (Tang et al. 2009).  For subsonic outflows the mixing is expected
  to be more local and more complete (Lu \& Wang 2011).  The magenta
  and cyan  ETGs in Fig.~\ref{f4} have gas densities and luminosities larger
than those considered by Tang et al. (2009); however, if a
discrete heating effect were still present at $\sigma_c<200$ km
s$^{-1}$, it could qualitatively explain a lower $f$ for these
galaxies.  There is also the possibility that $<T_*>$ has been
overestimated (and then the observed
$\Delta E_{th}$ underestimated) at the low
$\sigma_c$, if these ETGs are less pressure supported systems (e.g.,
Emsellem et al. 2011), and the stellar rotational streaming is not all thermalized
(Sect.~\ref{obs}). Another possible
explanation could be that ETGs with $\sigma_c>200$ km s$^{-1}$ are less
outflow-dominated than those at lower $\sigma_c$ (though this is not
supported just based on $L_X$, since magenta ETGs are found over the
whole
$\sigma_c$ range in Fig.~\ref{f4}), so that their
$E_{\rm grav}^-$ would be lower than assumed by the curves in
Fig.~\ref{f4}, and more SNIa's energy would be available for heating. In this
  way,  $f$ could have a value $<0.85$, possibly similar for all ETGs. 

\begin{figure}
\includegraphics[height=0.5\textheight,width=0.65\textwidth]{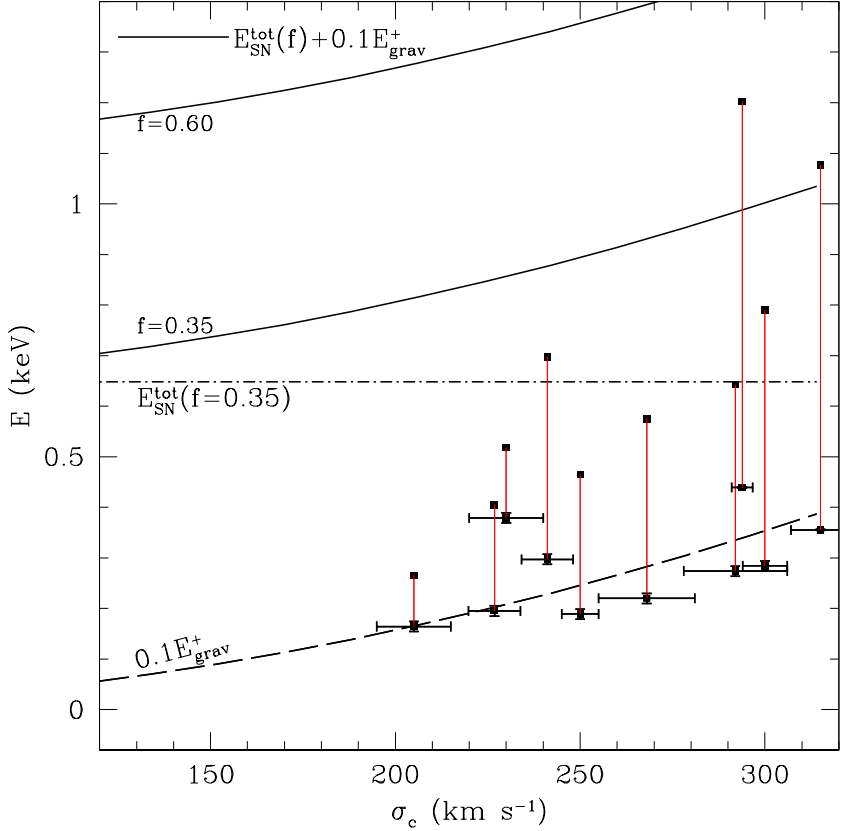}
\caption{The run with $\sigma_c$ of the energies provided by SNIa's
($E_{SN}^{tot}$) and gas infall ($E_{\rm grav}^+$); points with
errorbars show, for the ETGs with the largest $L_X$ in Fig.~\ref{f2},
the additional thermal energy ($\Delta E_{th}$) with respect to that
gained from the thermalization of the stellar random motions, required
to explain the observed $T$'s (see
Sect.~\ref{gravhea}); all these energies ($E_{SN}^{tot}$, $E_{\rm
  grav}^+$, and $\Delta E_{th}$), that are defined per unit mass in the
text, have been multiplied by $2\mu m_p/3$ to obtain their temperature-equivalent in
keV plotted here.  Solid lines show the sum of the energies provided by the SNIa's
and infall, for two cases of $f$, and for $E_{\rm grav}^+$ rescaled by
a factor of 0.1, and calculated for the galaxy mass model described by
the thick black line in Fig.~\ref{f2} (right panel).  The dashed line
gives for reference the value of $0.1 E_{\rm grav}^+$, and the
dot-dashed line the value of $E_{SN}^{tot}$ for $f=0.35$.  The points
with errorbars are obtained subtracting to the observed $kT$ the value
of $k<T_*>$ corresponding to its $\sigma_c$, for the mass model adopted for $E_{\rm
grav}^+$; each point is linked by a red line to a point including the
energy spent in radiation observed for that ETG (i.e., the upper point measures
$\Delta E_{th}+L_X/\dot M_*$).  See Sect.~\ref{gravhea} for more
details.}  \label{f3} \end{figure}

\begin{figure}
\includegraphics[height=0.5\textheight,width=0.65\textwidth]{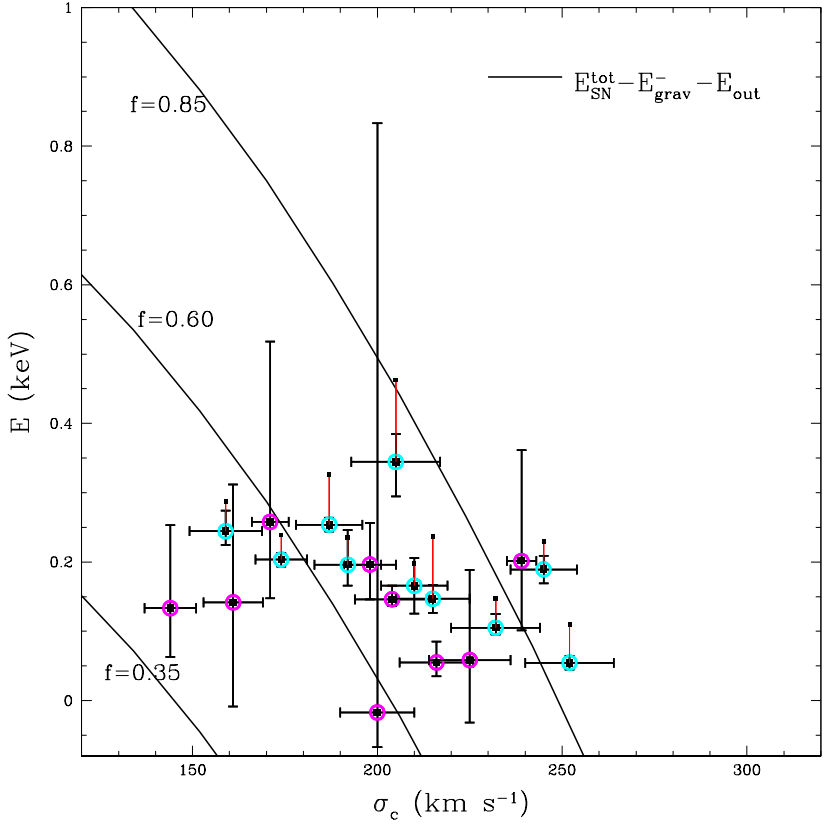}
\caption{
The run with $\sigma_c$ of the energy provided by SNIa's ($E_{SN}^{tot}$, for three
cases of $f$), after subtraction of the energy needed for the removal
of the gas from the galaxy ($E_{\rm grav}^-$), and for escape
($E_{out}$) at an average $v_{out}=c_s$(0.3 keV); all energies
have been computed as in Fig.~\ref{f3}. The adopted galaxy
mass model is that corresponding to the thick black line in
Fig.~\ref{f2} (right panel).  Points with errorbars show 
the additional thermal energy required to explain the observed $T$'s, 
calculated as in
Fig.~\ref{f3}, for ETGs with low/medium $L_X$ in Fig.~\ref{f2}.
For the cyan ETGs, a red line connects each point with the value including
the radiated energy, as in Fig.~\ref{f3}; the red lines 
of the magenta ETGs, whose $L_X$ values are the lowest, 
would be included within the colored circle, if shown.
See Sect.~\ref{sna} for more details.}
\label{f4}
\end{figure}

Finally, a comparison of Figs.~\ref{f3} and~\ref{f4} shows that the average
$\Delta E_{th}$ is slightly larger for the X-ray brightest ETGs (for
which it ranges between 0.1--0.5 keV) than for the X-ray faintest ones
(0--0.3 keV); moreover, while the $\Delta E_{th}$ in Fig.~\ref{f3}
can be explained even with $f<0.85$, it is required
that $f\sim 0.85$ for the X-ray faintest ETGs with the largest
$\sigma_c$ in Fig.~\ref{f4}. Both facts are the
consequence of the large fraction of the SNIa's energy input that is
used in gas extraction where outflows dominate, while all the SNIa's
energy remains within the galaxies where inflow dominates.

In conclusion, even for ETGs with low/medium $L_X$, a fundamental
X-ray property as $T$ can be accounted for by simple arguments, just
based on realistic galaxy mass models, and reasonable SNIa's heating
capabilities.  There may be, though, more energy available for
  the gas in ETGs with $\sigma_c<200$ km s$^{-1}$ than can be
  accounted for by the present simple scenario.

\subsection{The temperature and gas flows status in ETGs of
  intermediate mass}\label{reduc}

The observed X-ray properties (low $L_X$ and $T\sim <T_{\rm
  esc}^{sub}>$),  and the energy budget of the gas
(e.g., $<T_{\rm grav}^->$ vs. $<T_{inj}>$),
for ETGs with $\sigma_c \lsim 200$ km s$^{-1}$ are all consistent with
the expectations for outflows; the large $L_X$ and $<T_{\rm  grav}^->$ 
larger than $<T_{inj}>$ of ETGs
with $\sigma_c > 250$ km s$^{-1}$ agree with the gas being mostly
inflowing.  For $200<\sigma_c($km s$^{-1})<250$, instead, ETGs show
very different $L_X$ and $T$, whose values seem unrelated to
$<T_{\rm esc}^{sub}>$ and to the relative size of $<T_{\rm grav}^->$
  and $<T_{inj}>$ (Fig.~\ref{f2}). For example, for
$f=0.85$, $<T_{inj}>$ exceeds $<T_{\rm grav}^->$, but most $T$ values
lie well below $<T_{\rm esc}^{sub}>$, and even high $L_X$ values
(incompatible with outflows) are common. One
first explanation could be that $f<0.85$; for example, for
$f=0.35$, $<T_{inj}>$ becomes lower than $<T_{\rm grav}^->$ at
$\sigma_c\sim 180$ km s$^{-1}$ (Fig.~\ref{f2}).  Four ETGs with
$\sigma_c>200$ km s$^{-1}$ and a very low $L_X$ (Fig.~\ref{f2},
magenta symbols), though, require that $f>0.35$ in them, and then that
$f$ varies from galaxy to galaxy, or that their gas was removed by
other processes as an AGN outburst (e.g., Machacek et al. 2006, Ciotti
et al. 2010), or a merging or an interaction (Read \& Ponman 1998,
Sansom et al. 2006, Brassington et al. 2007). An event like the latter
two in the recent past is unlikely for three of these ETGs (NGC1023, NGC3115, NGC3379),
that are very regular in their stellar morphological and kinematic
properties , while in the other ETG (NGC4621) that hosts a counter-rotating core
(Wernli et al. 2002) it is possible.

A second explanation could be that $<T_{inj}>$ exceeds $<T_{\rm
  grav}^->$ only at the present epoch: while $<T_*>$ and $<T_{\rm
  grav}^->$ are independent of time, $T_{SN}$ may have been lower in
the past (Eq.~\ref{eq:tinj}), to the point that $<T_{inj}>$ may have
been lower than $<T_{\rm grav}^->$ for more ETGs than in Fig.~\ref{f2}
(that represents a snapshot of the present epoch). The gas then could
have accumulated and radiative losses have become important, even for
the gas injected in later epochs. In fact, the population synthesis
models of Sect.~\ref{heat} predict that $\dot M_*$ was larger at early
times (e.g., by $\sim 6$ times at an age of 3 Gyr), and then to keep
$T_{SN}$ high in the past, from Eq.~\ref{eq:tinj}, $\dot M_{SN}$ must
decrease with time $t$ at a rate similar to or steeper than that of
the stellar mass losses ($\dot M_*\propto t^{-1.3}$; Ciotti et
al. 1991).  Recent observational estimates indicate instead a SNIa's
rate decaying close to $t^{-1}$ (Maoz et al. 2010, Sharon et
al. 2010), thus $T_{SN}$ and $<T_{inj}>$ should be increasing with
time, reaching the values of Fig.~\ref{f2} at the present epoch. Then,
a ``cooling effect of the past'' would explain a moderate or high
$L_X$ even where $<T_{inj}>$ exceeds $<T_{\rm grav}^->$ in
Fig.~\ref{f2}.

Both explanations, $f<0.85$ and/or a lower $<T_{inj}>$ in the past, can
account for the lack of a widespread presence of outflows for
$200<\sigma_c$(km s$^{-1}) < 250$. Both, though, require a
mechanism different from the SNIa's energy input to cause degassing
in some low $L_X$ ETGs in the same $\sigma_c$ range.

Another possibility is that partial winds become common for
$\sigma_c>200$ km s$^{-1}$: these ETGs host an inner inflow and an
outer outflow (e.g., MacDonald \& Bailey 1978), with variations in the
galactic structure causing different sizes for the inflowing region,
and then different $L_X$ (Pellegrini \& Ciotti 1998). This possibility
holds even for $f=0.85$, and for both kinds of time evolution of
$<T_{inj}>$; in fact, if the flow is decoupled, ETGs may host a
central inflow even if $<T_{inj}>$ is larger than $<T_{\rm grav}^->$.
Similarly, the observed $T$'s can be lower than $<T_{\rm esc}^{sub}>$,
in ETGs where the outflow is only external; in this case, the observed
$T$'s may also be lowered by radiative losses in the central inflowing
region.

\section{The temperature of inflows and outflows}\label{temp}

In Fig.~\ref{f2}, the hottest gas is in ETGs with the highest $L_X$,
and the coolest one in ETGs with the lowest $L_X$ (as also found by BKF). This feature is
also present in the $L_X-\sigma_c$ relation, where ETGs with
$kT>0.4$ keV are the X-ray brightest (with one exception), while those
with $kT<0.3$ keV are the X-ray faintest (BKF).  All this may seem contrary
to the simple expectation that hotter gas is needed for escape, and
that the hotter the gas, the stronger the outflow, the lower the gas
content.  We re-examine below this point, first across the
whole $\sigma_c$ range, and then at fixed $\sigma_c$.

\begin{figure}
\includegraphics[height=0.5\textheight,width=0.65\textwidth]{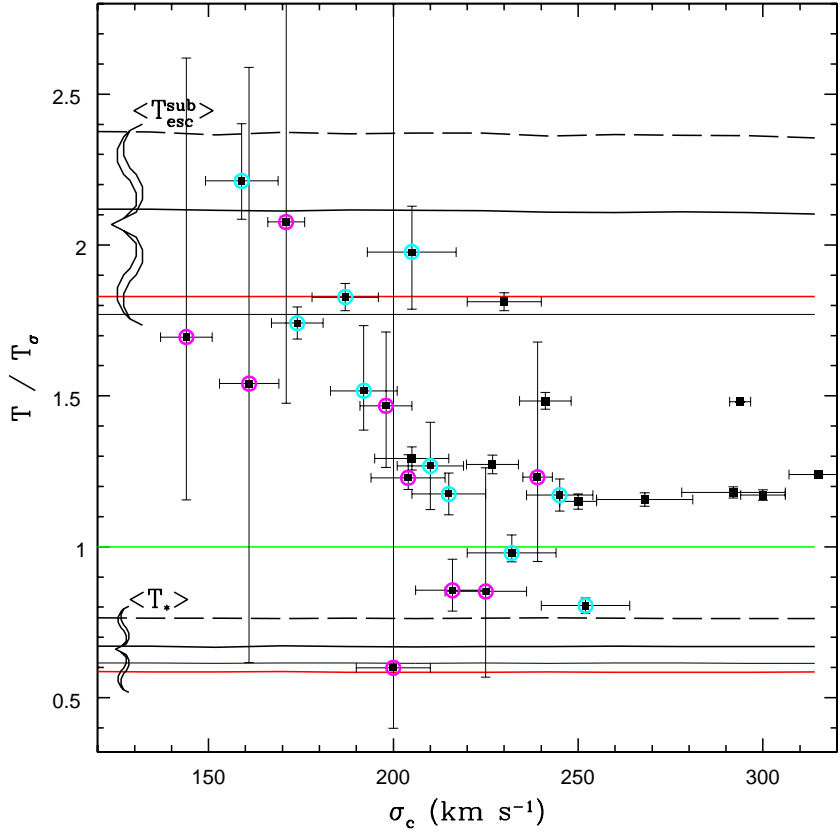}
\caption{The same as in Fig.~\ref{f2} (right panel), with  temperature values
rescaled by $T_{\sigma}$. }
\label{f5}
\end{figure}

A proper consideration of whether outflowing ETGs possess hotter or
colder gas than inflowing ones requires that all $T$'s are rescaled by
a temperature equivalent to the depth of the potential where the gas
resides (for example, by $T_{\sigma}$). Is there a trend then of the
distance of the observed $T$'s from $<T_{\rm esc}^{sub}>$, or $<T_*>$
?  This is examined by Fig.~\ref{f5}, where temperatures are rescaled
by $T_{\sigma}$, and the $\sigma_c^2$ dependency of all the curves in
Fig.~\ref{f2} is removed.  Figure~\ref{f5} shows that for $\sigma_c <
200$ km s$^{-1}$ the observed points reach $<T_{\rm
esc}^{sub}>/T_{\sigma}$, a result similar to that of Fig.~\ref{f2},
and that they fall below it with increasing $\sigma_c$ (with a
transition region of large dispersion in $T/T_{\sigma}$).  Therefore
ETGs with $\sigma_c < 200$ km s$^{-1}$ are indeed the hottest,
relatively to the virial temperature; since in these ETGs outflows are
important (Sect.~\ref{obs}), indeed the flow is relatively hotter in
outflows, and $T$ increases going from outflows to inflows only in an
absolute sense. The $T$'s of the X-ray brightest ETGs 
should show a dependence on
$\sigma_c^2$, if gravitational heating
of the gas dominates over SNIa's heating, and then they should
lie within a horizontal zone in Fig.~\ref{f5}. The observed distribution
does not disagree with this kind of dependence, but more cases are needed to
firmly establish its presence; such a dependence is not expected, though, since the
SNIa's heating should easily dominate over the gravitational one
(Sect.~\ref{gravhea}).

We finally compare the $T$'s at similar $\sigma_c$, in the most
populated region of Fig.~\ref{f5}, for $200<\sigma_c($km
s$^{-1})<250$. Here the variation of $T/T_{\sigma}$ is the largest,
and is covered by ETGs of all X-ray emission levels; the X-ray
brightest ETGs are found at $T/T_{\sigma}>1.1$, while the lowest
$T/T_{\sigma}$ values belong to the X-ray fainter ETGs.  For example,
two of the three lowest $T/T_{\sigma}$ values of the figure (those of
NGC3379 and NGC4621) belong to the X-ray faintest group. While
  heating sources seem abundant in gas-rich ETGs to account for their
  $T$'s (see, e.g., Fig.~\ref{f3}, and the additional possibility of
  MBH heating, Sect.~\ref{mbh}), even after taking into account their
  radiative losses, this result remains more difficult to explain for
ETGs of low/medium $L_X$, and may require ad hoc solutions.  It
  may be another representation of what mentioned in Sect.~\ref{sna},
  that $\Delta E_{th}$ can be larger for the X-ray brightest ETGs than
  for the X-ray faintest ones, due to the different employment of the
  SNIa's input energy; or it could be that $f < 0.85$ in these ETGs
so that SNIa's cannot make their gas hotter than this (Fig.~\ref{f4});
or their galaxy structure may be much different from an average one,
so that dividing all $T$'s for the same $T_{\sigma}$ produces a biased
view; or the evolutionary history of the gas may have been
peculiar. Certainly, this trend needs further investigation and, if
confirmed, it will provide the basis for further theoretical work.

\section{Conclusions}\label{concl}

This work has focussed on the origin of the hot gas temperatures
recently derived for a sample of ETGs observed with $Chandra$ down to
galaxy masses and X-ray luminosities smaller than ever before.  A few
characteristic mass-weighted average temperatures have been defined for a gas
distribution $\rho_{gas}(r)\propto \rho_*(r)$, as for the gas shed by
stars: the virial temperature $<T_*>$; the injection temperature
$<T_{inj}>$, as the sum of $<T_*>$ and of a temperature equivalent to
the SNIa's kinetic energy input (with a factor $f$ allowing for its
uncertain thermalization); the escape temperature $<T_{\rm grav}^->$,
defined as the temperature equivalent of the energy required for
escape from the gravitational potential; a fiducial value
for the temperature of escaping gas, evaluated on a streamline of very
subsonic velocity  ($<T_{\rm esc}^{sub}>=0.6 <T_{\rm grav}^->$); and
finally, the temperature equivalent to the energy liberated by the gas
inflow to the galactic center, $<T_{\rm grav}^+>$. These
temperatures were then calculated for a set of representative galaxy
mass models, made by the superposition of a central MBH, and a stellar
and a dark mass density distributions, with parameters constrained
from the fundamental scaling laws of ETGs and recent observational
findings.  The main properties of the characteristic temperatures are
that:

$\bullet$ All temperatures scale as $\sigma_c^2$ (except for
$<T_{inj}>$), and increase for larger and/or more concentrated mass
content.  For the adopted set of representative galaxy mass models,
$<T_*>$ is lower than $T_{\sigma}$ (by $\sim 0.1-0.2$ keV), $<T_{\rm
  grav}^->\approx 5 <T_*>$, and $<T_{\rm grav}^+>\approx 2<T_{\rm
  grav}^->$; the temperature that can be produced by infall heating,
though, will be much lower than $<T_{\rm grav}^->$, due to
energy losses in radiation, kinetic energy of mass condensations, and
mass drop-outs from the flow.

$\bullet$ $<T_{inj}>$ is by far the largest of the characteristic
temperatures, due to the important SNIa's contribution (independent of
$\sigma_c$); for $f=0.85$, it is larger than the minimum injection
temperature for global escape up to $\sigma_c\sim 250 $ km s$^{-1}$.

The comparison of the characteristic temperatures with those observed,
in the $T-\sigma_c$ plane, shows that:

$\bullet$ The best fit $T-\sigma_c$ relation previously found for
X-ray bright ETGs reproduces the average trend of the observed $T$
down to low temperatures, and low $L_X$. ETGs with low/medium $L_X$
show the largest departures from this fit, which can be explained by
the variety of gas flow phases possible in them (winds, subsonic
outflows, partial winds), where the main input energies (from SNIa's and gas
infall) are used in different ways.

$\bullet$ All observed $T$'s are larger than $<T_*>$; the additional
heating of the gas $\Delta E_{th}$, with respect to that provided by 
the thermalization
of the stellar motions, is $\Delta E_{th}\approx 0-0.3$ keV for the
X-ray faintest ETGs, and $\Delta E_{th}\approx 0.1-0.5$ keV for the
X-ray brightest (for a representative galaxy mass model).

$\bullet$ In a stationary situation, $\Delta E_{th}$ of the X-ray
brightest ETGs can be accounted for by the energy input of SNIa's and gas
infall, even if they are much reduced with respect to standard
assumptions (i.e., $f$ can be $<0.85$). The gravitational heating
produces a $T\propto \sigma_c^2$ trend,
that may be present in the X-ray brightest ETGs; the SNIa's
heating, though, is expected to be dominant.

$\bullet$ $\Delta E_{th}$ can be provided by SNIa's in X-ray fainter
ETGs, where outflows are important; most of the SNIa's energy is
needed for gas extraction, and less for the kinetic energy of the
escape.  The value of $f$ to account for the observed $\Delta E_{th}$
increases with $\sigma_c$, until the whole SNIa's energy ($f\approx
0.85$) is required at the highest $\sigma_c$. With this $f$,
  though, at low $\sigma_c$ the observed $\Delta E_{th}$ are lower
  than expected. Possible solutions require a different efficiency of
  the SNIa's energy mixing process, or an overestimate of $<T_*>$ at
  low $\sigma_c$ if these ETGs are less pressure supported systems, or
  a more complex flow status than in the simple scheme adopted.

$\bullet$ At low $\sigma_c \lsim 200$ km s$^{-1}$, $<T_{inj}>$ is
larger than $<T_{\rm grav}^->$, the $L_X$ values are low
and the $T$'s are of the order of $<T_{\rm esc}^{sub}>$: all this 
agrees well with what expected for
outflows. At high $\sigma_c > 250$ km s$^{-1}$, $<T_{inj}>$ is
lower than $<T_{\rm grav}^->$, and the high $L_X$ and $T$ can be explained 
with the gas mostly inflowing.  For $200<\sigma_c($km s$^{-1})<250$, instead,
there is a large variation in $L_X$ and $T$. Possible explanations
could be that the SNIa's energy input varies from galaxy to galaxy,
and/or that $<T_{inj}>$ was lower in the past, due to the different
time evolution of the mass loss and the SNIa's rate; or
that partial winds become common, with the flow status less related to
the values of $<T_{inj}>$, $<T_{\rm grav}^->$, and $<T_{\rm esc}^{sub}>$.

$\bullet$ When measured relatively to the depth of the potential well,
the observed temperatures $T/T_{\sigma}$ are larger for $\sigma_c < 200 $ km
s$^{-1}$ (outflows), and lower for $\sigma_c>250$ km s$^{-1}$
(inflows).  The observed $T$'s then increase from outflows to inflows
only in an absolute sense, and the gas is relatively hotter in
outflows. In the intermediate region of $200<\sigma_c($km
s$^{-1})<250$, lower $L_X$ values tend to correspond to lower $T$ and
$T/T_{\sigma}$, which requires ad hoc explanations, and then deserves
further observational and theoretical investigation.

\acknowledgments I thank Luca Ciotti for helpful discussions, and
Dong-Woo Kim and the referee for useful comments.

\clearpage

\begin{table}
\caption[] { Observed properties of the ETG sample.}\label{tab1}
\begin{tabular}{ l  c c c c l }
\noalign{\smallskip}
\hline
\noalign{\smallskip}
 Name   & $kT$    & 1$\sigma$ error & $L_X$                     & $\sigma_c$ &  Ref \\
             & (keV)    &                            & ($10^{40}$ erg             s$^{-1}$)  & (km s$^{-1})$  & \\
   (1)      &   (2)      &         (3)   &                (4)   &     (5)         &   (6)        \\
\noalign{\smallskip}
\hline
\noalign{\smallskip}
NGC 720 &   0.54 & -0.01; +0.01 &   5.06 &     241 & Binney et
al. 1990 \\
NGC 821 &   0.15 & -0.05; +0.85 &   $2.13\times 10^{-3}$  &     200 & Kuntschner et al. 2010\\
NGC1023 &   0.32 & -0.01; +0.02 &   $6.25\times 10^{-2}$ &     204 &Kuntschner et al. 2010\\
NGC1052 &   0.34 & -0.02; +0.02 &   $4.37\times 10^{-1}$ &     215 &Binney et
al. 1990\\ 
NGC1316 &   0.60 & -0.01; +0.01 &   5.35 &     230 &D'Onofrio et
al. 1995 \\ 
NGC1427 &   0.38 & -0.11; +0.26 &   $5.94\times 10^{-2}$ &     171 &D'Onofrio et
al. 1995\\ 
NGC1549 &   0.35 & -0.04; +0.04 &   $3.08\times 10^{-1}$ &     210 &Longo et al. 1994\\ 
NGC2434 &   0.52 & -0.05; +0.04 &   $7.56\times 10^{-1}$ &     205 &Longo et al. 1994\\ 
NGC2768 &   0.34 & -0.01; +0.01 &   1.26 &     205 &Kuntschner et al. 2010 \\ 
NGC3115 &   0.44 & -0.10; +0.16 &   $2.51\times 10^{-2}$ &     239 &Fisher 1997\\ 
NGC3377 &   0.22 & -0.07; +0.12 &   $1.17\times 10^{-2}$ &     144 &Kuntschner et al. 2010\\ 
NGC3379 &   0.25 & -0.02; +0.03 &   $4.69\times 10^{-2}$ &     216 &Kuntschner et al. 2010\\ 
NGC3384 &   0.25 & -0.15; +0.17 &   $3.50\times 10^{-2}$ &     161 &Kuntschner et al. 2010\\ 
NGC3585 &   0.36 & -0.05; +0.06 &   $1.47\times 10^{-1}$ &     198 &Fisher 1997\\ 
NGC3923 &   0.45 & -0.01; +0.01 &   4.41 &     250 &Pellegrini et
al. 1997\\ 
NGC4125 &   0.41 & -0.01; +0.01 &   3.18 &     227 &Bender et
al. 1994\\ 
NGC4261 &   0.66 & -0.01; +0.01 &   7.02 &     300 &Bender et
al. 1994\\ 
NGC4278 &   0.32 & -0.01; +0.01 &   $2.63\times 10^{-1}$ &     252 &Kuntschner et al. 2010\\ 
NGC4365 &   0.44 & -0.02; +0.02 &   $5.12\times 10^{-1}$ &     245 &Bender et
al. 1994\\ 
NGC4374 &   0.63 & -0.01; +0.01 &   5.95 &     292 &Kuntschner et al. 2010 \\ 
NGC4382 &   0.40 & -0.01; +0.01 &   1.19 &     187 &Kuntschner et al. 2010\\ 
NGC4472 &   0.80 & -0.00; +0.00 &   18.9 &     294 &Bender et
al. 1994\\ 
NGC4473 &   0.35 & -0.03; +0.05 &   $1.85\times 10^{-1}$ &     192 &Kuntschner et al. 2010\\ 
NGC4526 &   0.33 & -0.01; +0.02 &   $3.28\times 10^{-1}$ &     232 &Kuntschner et al. 2010\\ 
NGC4552 &   0.52 & -0.01; +0.01 &   2.31 &     268 &Kuntschner et al. 2010 \\ 
NGC4621 &   0.27 & -0.09; +0.13 &   $6.08\times 10^{-2}$ &     225 &Kuntschner et al. 2010\\ 
NGC4649 &   0.77 & -0.00; +0.00 &   11.7 &     315 &Bender et al.
1994\\ 
NGC4697 &   0.33 & -0.01; +0.01 &   $1.91\times 10^{-1}$ &     174 &Binney et
al. 1990\\ 
NGC5866 &   0.35 & -0.02; +0.03 &   $2.42\times 10^{-1}$ &     159 & Fisher 1997\\ 
\hline
\end{tabular}

Column (1): galaxy name. Cols. (2), (3) and (4): the hot gas temperature,
its uncertainty, and the 0.3--8 keV gas luminosity, from
BKF. Col. (5): the stellar velocity dispersion, as the
luminosity-weighted average within an aperture of radius $R_e/8$, with
its reference in Col. (6).

\end{table}

\clearpage
\begin{figure}
\includegraphics[height=0.55\textheight,width=0.85\textwidth]{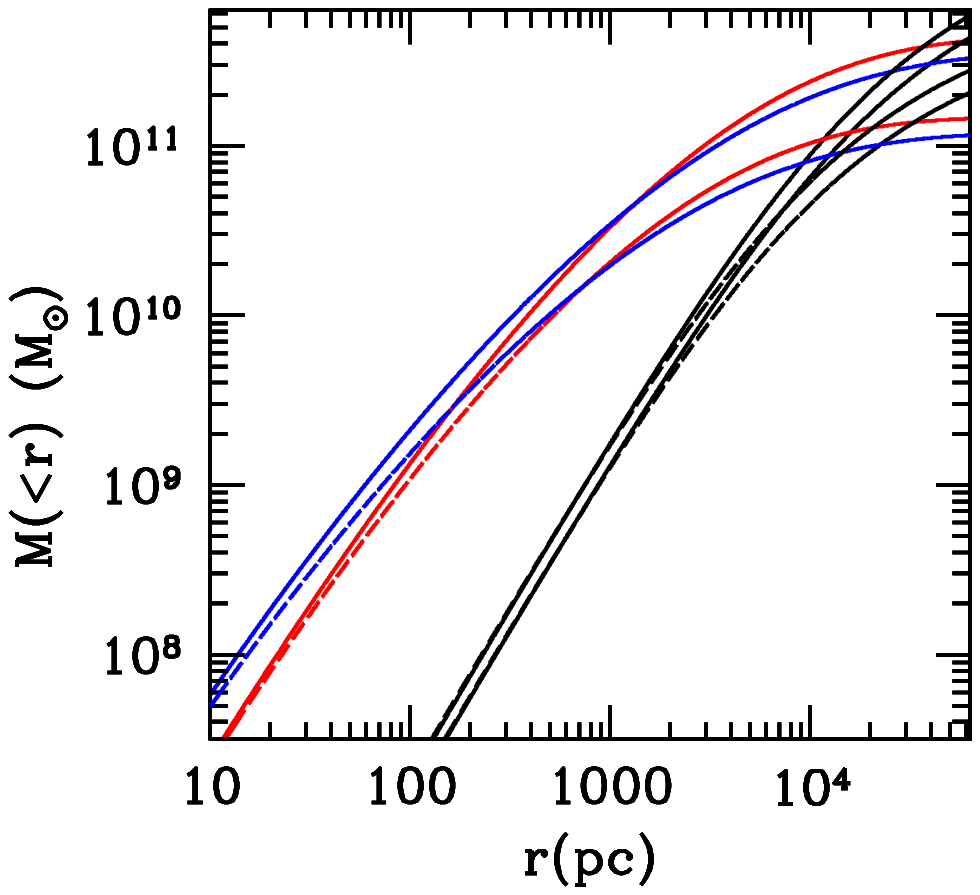}
\vskip -7.6truecm
\includegraphics[height=0.55\textheight,width=0.85\textwidth]{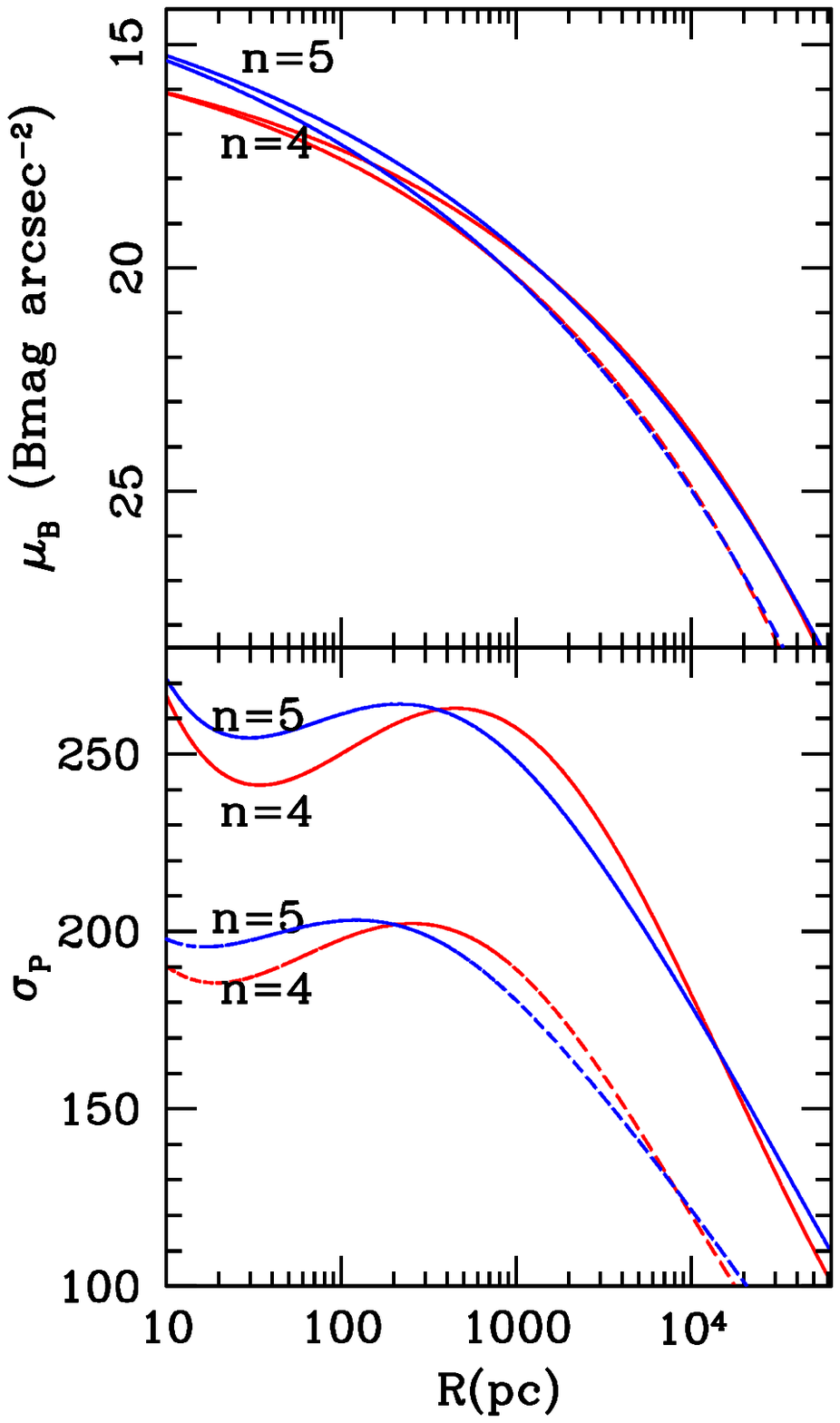}
\caption{Mass (up), B-band surface brightness (middle), and projected
velocity dispersion (bottom) profiles of three-component galaxy models
(MBH$+$stars$+$dark matter), for two representative ETGs with
isotropic orbits and an aperture velocity dispersion within $R_e/8$ of
$\sigma_c = 260$ km s$^{-1}$ (solid lines; $L_B=5\times
10^{10}L_{B,\odot}$, and $R_e=6.5$ kpc), and of $\sigma_c = 200$ km
s$^{-1}$ (dashed lines; $L_B=2\times 10^{10}L_{B,\odot}$, and
$R_e=3.6$ kpc).  Red lines refer to a stellar S\'ersic profile with
index $n=4$, blue ones with $n=5$. The dark halo in the upper panel
(black, with the same line type as the corresponding stellar profile) 
follows the NFW profile, with $\beta = 2$, and ${\cal R}=3$ (for $n=4$) or
${\cal R}= 5$ (for $n=5$), and then ${\cal R}_e$=0.24 or 0.41, from the
Jeans equations (see Sect.~\ref{mass}).}
\label{f1} 
\end{figure}

\clearpage

\clearpage

\clearpage

\end{document}